# Brain-Like Infrastructure for Embedded SoC Diagnosis

V. I. Hahanov[1], Wajeb Gharibi[2], Olesya Guz[3]

[1] Computer Engineering Faculty, Kharkov National University of Radioelectronics, Kharkov, Ukraine
hahanov@kture.kharkov.ua
[2] Kingdom of Saudi Arabia, Jazan University, gharibiw2002@yahoo.com
[3] Road Transport Faculty, Donetsk Institute of Road Transport, Donetsk, Ukraine kiu@kture.kharkov.ua

*Abstract*- **This article describes high-speed multiprocessor architecture for the concurrent analyzing information represented in analytic, graph- and table forms of associative relations to search, recognize and make a decision in n-dimensional vector discrete space. Vector-logical process models of actual applications, for which the quality of solution is estimated by the proposed integral non-arithmetical metric of the interaction between Boolean vectors, are described.**

## I. INTRODUCTION

The aim is to remove arithmetic from computer and transform free resources to the brain-like infrastructure of associative logic simulating the brain functionality that makes possible making the right decision every moment. The brain and the computer have the same technological basis in the form of primitive logical operations: and, or, not, xor. With experience, the brain and the computer create more complex functional space-time logic converters using the above primitive operations. Specialization of computer, focused on using only logical operations, enables to approximate to the associative logic human thinking, and thus considerably (x100) improve the performance of solving nonarithmetic problems.

Removing arithmetic operations, leveraging the parallelism of the vector logic algebra, and multiprocessor architecture provide an efficient infrastructure, which combines mathematical and technological culture to solve applied problems.

Brain-likeness of multiprocessor digital system-on-a-chip is the concept of making an architecture and models of computational processes to implement typical brain nonarithmetic associative logic functionalities on today's digital platform by using vector logical operations and criteria for search, pattern recognition and decision-making problems. Market appeal of logical associative multiprocessor (LAMP) is determined by thousands of old and new logical problems, which now are solved ineffectively by redundant universal computers with high-performance arithmetic processor. Here are some problems relevant to the IT-market: 1. Analysis and synthesis of syntactic and semantic language structures (abstracting, error correction, analysis of the text quality). 2. Video and audio pattern recognition by means of their representation by vector models of essential parameters in discrete space. 3. Use of Infrastructure IP for complex technical products to ensure their manufacturability and lifetime reliability. 4. Knowledge testing and expert appraisal of objects or parties to determine their validity. 5. Identification of the object or process to make a decision under uncertainty. 6. Exact information retrieval in the Internet, if information is given by a vector of parameters. 7. Target designation of fighter or aircraft autoland system functioning in microsecond time. 8. Air traffic control or optimization of municipal traffic control infrastructure to avoid conflicts. Practically all these problems are solved in real time; they are isomorphic by the logical structure of the process models, based on a set of interrelated associative tables. To solve them it is necessary quick and dedicated hardware platform (LAMP), focused on the concurrent execution of search, recognition and decision-making procedures, estimated by means of the integral nonarithmetic quality criterion.

Our goal in this article is to increase considerably (x100) the speed of search, recognition and decision-making procedures by means of multiprocessor and concurrent implementation of associative logic vector operations for the analyzing graph and tabular data structures in discrete Boolean space without the use of arithmetic operations.

The problems: 1) To develop nonarithmetic metric for estimating the associative logic solutions. 2) To create data structures and process models for solving the applied problems. 3) To design architecture of logical associative multiprocessor. 4) To implement LAMP.

Essence of the research is the infrastructure for expert servicing of requests in real time integrating multiprocessor system-on-chip with associative-logical data structures to obtain a deterministic solution, the validity of which is estimated by nonarithmetic integral interaction quality criterion of a query and given discrete space.

Object of the research is the infrastructure of search, recognition and decision-making in the discrete Boolean space based on vector logic algebra, multiprocessor platform for the analyzing associative logical data structures, and non-arithmetic integral quality criterion.

Subject of the research is associative logical data structures and process models for searching, recognizing and decision-making, based on nonarithmetic integral quality criterion by using multiprocessor system-on-a-chip, focused on vector logical operations.

References: 1. Hardware platform for associative logical information analysis [1-2]. 2. Associative logical data structures for solving the information problems [3-4]. 3. Models and methods for discrete analyzing and synthesizing [5-6]. 4. Mul-tiprocessors for solving information-logical problems [7-10]. 5. Brain-like and intelligent logical computing [11-12].

## II. INTEGRAL METRIC FOR SOLUTION ESTIMATION

Infrastructure of brain-like multiprocessor includes models, methods and associative logical data structure, focused on hardware support of search, recognition and decision-making processes [22-24] on the basis of vector nonarithmetic operations.

Evaluation of problem solution is determined by the vector-logical criterion of interaction quality between a query (a vector m) and a system of associative vectors (associators). The query processing results in generating a positive response in the form of one or more associators, as well as the numerical grade of membership characteristic (quality function) of an input vector m to the obtained solution: $\mu(m \in A)$. The input vector $m = (m_1, m_2, ..., m_i, ..., m_q)$, $m_i \in \{0,1,x\}$ and the matrix $A_i$ of associators $A_{ijr} (\in A_{ij} \in A_i \in A) = \{0,1,x\}$ have the same dimension that is equal to $q$. Below the membership grade of m-vector to A is designated by $\mu(m \in A)$.

There are 5 types of set-theoretic (logical) $\Delta$ - interaction of two vectors $m \cap A$ defined in Fig. 1. They form all primitive reactions of the generalized SRM systems (SRM – Search, Pattern Recognition and Decision Making) on the input request vector. In the technological field of knowledge – Design & Test – this sequence of actions is isomorphic to the route: fault finding, fault locating, decision-making for repairing. All three stages of technological route require the metric for estimating solutions to choose the optimal variant.

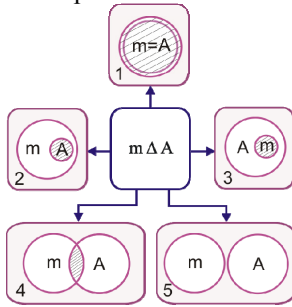

Fig. 1. The results of the intersection of two vectors

Definition. Integral set-theoretic metric for the estimating query quality is a function of the interaction of multivalued vectors $m \cap A$, which is determined by average sum of three normalized parameters: code distance $d(m,A)$, membership function $\mu(m \in A)$ and membership function $\mu(A \in m)$:

$$Q = \frac{1}{3}[d(m, A) + \mu(m \in A) + \mu(A \in m)],$$
$$d(m, A) = \frac{1}{n}[n - card(m_i \bigcap_{i=1}^{n} A_i = \emptyset)];$$
$$\mu(m \in A) = 2^{card(m \cap A) - card(A)} \leftarrow card(m \cap A) =$$
$$= card(m_i \bigcap_{i=1}^{n} A_i = x) \,\&\, card(A) = card(\bigcup_{i=1}^{n} A_i = x); \quad (1)$$
$$\mu(A \in m) = 2^{card(m \cap A) - card(m)} \leftarrow card(m \cap A) =$$
$$= card(m_i \bigcap_{i=1}^{n} A_i = x) \,\&\, card(m) = card(\bigcup_{i=1}^{n} m_i = x).$$

Explanations. The normalization of parameters makes it possible to estimate the level of vector interaction in the interval [0,1]. If it is fixed the limiting maximum value of each parameter equal to 1, it means the vectors are equal. The minimal estimation $Q = 0$ is fixed if the vectors are not coincided by all n coordinates. If intersection power $m \cap A = m$ is equal to half of $A$ vector space, membership and quality functions are equal respectively:

$$\mu(m \in A) = \frac{1}{2}; \,\mu(A \in m) = 1; \,d(m, A) = 1; \,Q(m, A) = \frac{5}{2 \times 3} = \frac{5}{6}.$$

The same value will be setting for Q parameter if the power of intersection $m \cap A = A$ is equal to half of the vector space $m$. If the power of intersection $card(m \cap A)$ is equal to half of the power of vector spaces $A$ and $m$, membership functions are the following:

$$\mu(m \in A) = \frac{1}{2}; \,\mu(A \in m) = \frac{1}{2}; \,d(m, A) = 1; Q(m, A) = \frac{4}{2 \times 3} = \frac{2}{3}.$$

It should be noted, if the intersection of two vectors is equal to the empty set, then the power of number 2 from the symbol "empty" is equal to zero: $2^{card(m \cap A) = \emptyset} = 2^{\emptyset} = 0$. It really means that the number of common points in the intersection of two spaces is zero.

The aim of a new vector logical criterion of solution quality is improving considerably the performance of calculating the quality Q of interaction between the components m and A, when analyzing the associative data structures by using the vector logical operations only. The arithmetic criterion (1) without the averaging membership functions and code distance can be transformed to the form:

$$Q = d[m, A_{i(j)}] + \mu[m \in A_{i(j)}] + \mu[A_{i(j)} \in m],$$
$$d(m, A_{i(j)}) = card[m \underset{i(j)=1}{\overset{n(m)}{\oplus}} A_{i(j)} = 1];$$
$$\mu(m \in A_{i(j)}) = card[A_{i(j)} = 1] - card[m \underset{i(j)=1}{\overset{n(m)}{\wedge}} A_{i(j)} = 1]; \quad (2)$$
$$\mu(A_{i(j)} \in m) = card[m = 1] - card[m \underset{i(j)=1}{\overset{n(m)}{\wedge}} A_{i(j)} = 1].$$

The first component of the criterion forms the degree of mismatch between n-dimensional vectors – the code distance, by performing xor operation, second and third ones determine the degree of non-membership of conjunction result to a set of "1" each of two interacting vectors. The notions of membership and non-membership are complementary, but calculating non-membership is more technological. Thus, the ideal criterion of quality is equal to zero, if two vectors are equal. The estimation of the interaction quality between two binary vectors is decreasing with increasing criterion from 0 up to 1. To finally remove arithmetic operations, when counting a vector quality criterion, it is necessary to transform the expressions (2) to the form:

$$Q = d(m,A) \vee \mu(m \in A) \vee \mu(A \in m),$$
$$d(m,A) = m \oplus A;$$
$$\mu(m \in A) = A \wedge \overline{m \wedge A};$$
$$\mu(A \in m) = m \wedge \overline{m \wedge A}.$$
(3)

Here the criteria are not numbers, but vectors, which determine the interaction of components $m, A$. The increasing quantity of 0 in three quality vectors improves the criterion, and 1's indicate loss of interaction quality. To compare the estimations it is necessary to determine the power of 1's in each vector without performing addition operation. This can be done using the register [4] (Fig. 2), which makes it possible to perform left shifting and compacting all 1 coordinates of n-bit binary vector for one clock cycle.

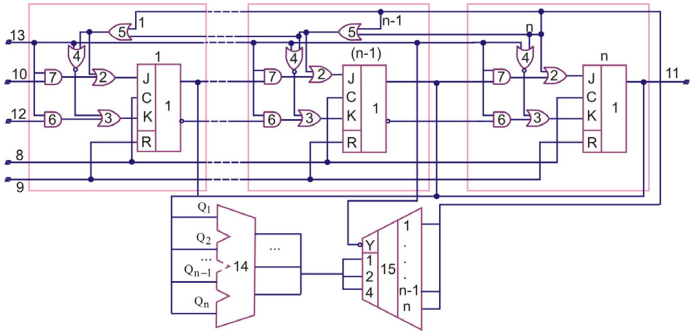

Fig. 2. Register for shifting and compacting 1's

After compacting procedure right unit bit number of compacted set of 1's determines the index of interaction quality for vectors. For binary sets $m = (110011001100)$, $A = (000011110101)$ the determining their interaction quality by formulas (3) is shown in the following form (zero coordinates are marked by dots):

| | | 
|---|---|
| m | 1 1 . . 1 1 . . 1 1 . . |
| A | . . . . 1 1 1 1 . 1 . 1 |
| $m \wedge A$ | . . . . 1 1 . . . 1 . . |
| $\overline{m \wedge A}$ | 1 1 1 1 . . 1 1 1 . 1 1 |
| $d(m,A) = m \oplus A$ | 1 1 . . . . 1 1 1 . . 1 |
| $\mu(A \in m) = m \wedge \overline{m \wedge A}$ | 1 1 . . . . . . 1 . . . |
| $\mu(m \in A) = A \wedge \overline{m \wedge A}$ | . . . . . . 1 1 . . . 1 |
| $Q = d(m,A) \vee \mu(m \in A) \vee \mu(A \in m)$ | 1 1 . . . . 1 1 1 . . 1 |
| $Q(m,A) = (6/12)$ | 1 1 1 1 1 1 . . . . . . |

It is formed not only the estimation of vector interaction that is equal to $Q = d(m,A) \vee \mu(m \in A) \vee \mu(A \in m)$, but the most importantly, the unit row coordinates identify all essential variables for which there is low-quality vector interaction. To compare two solutions obtained by logical analysis, compressed quality vectors Q are used; and vector procedure including the following vector operations is performed:

$$Q(m,A) = \begin{cases} Q_1(m,A) \leftarrow or[Q_1(m,A) \wedge Q_2(m,A) \oplus Q_1(m,A)] = 0; \\ Q_2(m,A) \leftarrow or[Q_1(m,A) \wedge Q_2(m,A) \oplus Q_1(m,A)] = 1. \end{cases}$$ (4)

Vector-bit or-operator of devectorization determines a binary bit solution on the basis of application a logical OR operation to n bits of an essential variables vector of quality criterion. A circuit design for decision

$$Q = \begin{cases} Q_1 \leftarrow Y = 0 \\ Q_2 \leftarrow Y = 1 \end{cases}$$

and analytic process-model include three operations, shown in Fig. 3.

| $Y = \vee[(Q_1 \wedge Q_2) \oplus Q_1]$ | 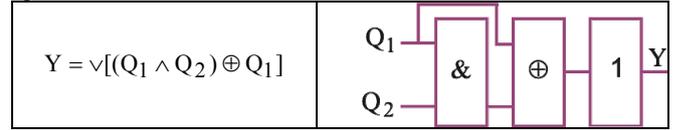 |
|---|---|

Fig. 3. Process-model of decision

For binary vectors which are quality criteria the procedure for choosing the best one on the basis of expression (4) is presented below:

| $Q_1(m,A) = (6,12)$ | 1 1 1 1 1 1 . . . . . . |
|---|---|
| $Q_2(m,A) = (8,12)$ | 1 1 1 1 1 1 1 1 . . . . |
| $Q_1(m,A) \wedge Q_2(m,A)$ | 1 1 1 1 1 1 . . . . . . |
| $Q_1(m,A) \oplus Q_1(m,A) \wedge Q_2(m,A)$ | . . . . . . . . . . . . |
| $Q(m,A) = Q_1(m,A)$ | 1 1 1 1 1 1 . . . . . . |

Vector logical criteria of interaction quality for associative sets enable to obtain estimation of the search, pattern recognition and decision-making with high-speed parallel logic operations, which is especially important for critical real-time systems.

III. ARCHITECTURE OF LOGIC ASSOCIATIVE MULTIPROCESSOR

To analyze large information volumes of logical data, there are several technologies focused to the practical application: 1. Using a workstation for serial programming, where the cost and time of problem solving are very high. 2. Development of a dedicated concurrent processor based on the PLD. The high concurrency of information processing compensates for the relatively low clock rate in comparison with CPU. Such reprogrammable circuit design is the best solution regarding performance. Disadvantage is lack of flexibility the software methods for solving logic problems and high cost of implementing the system-on-a-chip PLD under large production volumes. 3. The best solution is to leverage advantages CPU, PLD and ASIC concurrently [15,16]. This is due to the flexibility of programming, the possibility of correcting the source code, the minimum command set, and simple circuit designs for hardware multiprocessor implementation, the parallelization of logic procedures by the structure of bit processors. The implementation of a multiprocessor in ASIC allows to obtain the maximum clock rate, the minimum chip cost for large product volumes, and low power consumption. Combining the advantages of the technologies above determines the basic configuration of the LAMP, which has spherical multiprocessor structure (Fig. 3), consisting of 16 vector sequencers. Each sequencer together with the boundary elements is connected with eight contiguous ones. The processor PRUS [17], developed by Dr. Stanley Hyduke (CEO Aldec, USA), is the LAMP prototype.

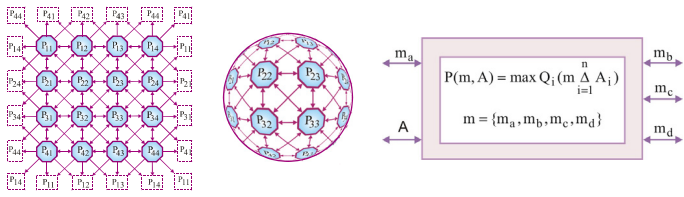

Fig. 3. LAMP macroarchitecture and interface

Entering information in the processor is realized like the classical design flow, except the stage "place and route" that is replaced by the operation of distributing software modules and data among all logical bit processors running concurrently. The compiler provides the placement of data among processors, sets the time of searching for solutions at the output each of them, and also plans transfer the results to another processor. LAMP is an effective processor network, which processes the data and provides the exchange of information between network components when searching for solution. The simple circuit engineering of each processor can effectively process very large arrays with millions bits of information, expending time in hundreds the times less compared with general-purpose processor. Basic cell (vector processor for LAMP) can be synthesized by using 200 gates, which makes it possible to implement network containing 4096 computers in ASIC, using advanced silicone technology. Taking into account that memory costs for data storage are very small, LAMP may be applied for the designing of control systems in the areas of human activity, such as: industry, medicine, information protection, geology, weather forecasting, artificial intelligence, space science. LAMP is of particular interest for digital data processing, pattern recognition and cryptanalysis. If LAMP functioning is considered, its main purpose is obtaining quasi-optimal solution of the integrated problem of search and / or pattern recognition by using infrastructure components focused to the performing vector logical operations:

$$P(m, A) = \min_i Q_i(m \underset{i=1}{\overset{n}{\Delta}} A_i), m = \{m_a, m_b, m_c, m_d\}.$$ System interface, corresponding to this functional, is presented in Fig. 4. All components $\{A, m_a, m_b, m_c, m_d\}$ can be input and output. Bidirectional interface specification is related to the invariance of relation for all variables, vectors, A-matrix, components and infrastructure inputs and / or outputs. Therefore, the structural model of LAMP can be used to solve any problems of direct and inverse implication in discrete logical space, and it emphasizes its difference from the automaton model concept of computer with explicit inputs and outputs. The components or registers $m = (m_a, m_b, m_c, m_d)$ are used for solution in the form of buffer, input and output vectors, as well as for identification of quality estimation for request performance. One of the variants multiprocessor architecture LAMP is a structure shown in Fig. 4. The main its component is a multiprocessor matrix $P = [P_{ij}], card(4 \times 4)$, containing 16 vector-processors, each of them is designed for performing 5 logic vector operations with data memory contents, described by a table of dimension $A = card(m \times n)$.

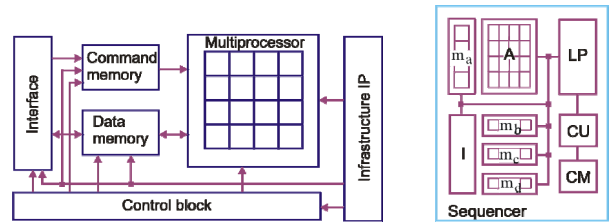

Fig. 4. LAMP architecture and sequencer structure

Interface is used for data exchange and data loading to the appropriate memory commands. The control unit initializes the executing commands of logical data processing and synchronizes the functioning all components of a multiprocessor. Infrastructure IP [1] is designed for servicing all modules, diagnosing faults and repairing functionality of components and device in whole. Elementary logic associative processor or sequencer (see Fig. 4) is a part of the multiprocessor and contains: logical processor (LP), associative (memory) A-matrix for concurrent executing basic operations, block of vectors m, designed for concurrent processing rows and columns of A-matrix, as well as data exchange when computing, direct access memory (CM) for the storing commands of data processing software, automaton (CU) for logic operations execution control, interface (I) for the connecting sequencer and other elements of a multiprocessor. Logical Processor (LP) (Fig. 5) provides the implementation of five operations (and, or, not, xor, s1s - shift left bit crowding), which are the basis for the creating algorithms and procedures of information retrieval and evaluation of solutions. LP module has a multiplexer at the input to select one of five operands, which is passed to the selected logic vector operator. By using a multiplexer (element or), a result is entered in one of four operands, which are selected by appropriate address.

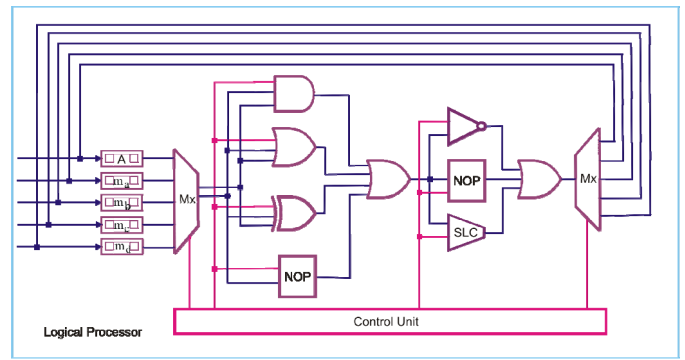

Fig. 5. Structure of logic calculations

Implementation features of the logical processor is use of three binary (and, or, xor) and two unary (not, slc) operations. The last ones can be added to the cycle of processing register data by selecting one of three operations (not, slc, nop – no-operation). To improve the efficiency of logical unit, two ele-ments with empty operation are included. If it is necessary to perform a unary operation only, the selecting nop at the level of binary commands should be done, that almost means the transfer data through a follower to second level of unary operations. All LP operations are register or register-matrix. The last ones are designed for the analyzing vector-rows of a table using input m-vector as a request for exact information retrieval. The following combination of operators and operands are acceptable in a unit for logic calculating:

$$C = \begin{cases} \{m_a, m_b, m_c, m_d\} \Delta A_i; \\ \{m_a, m_b, m_c, m_d\} \Delta \{m_a, m_b, m_c, m_d\}; \\ \{not, nop, slc\} \{m_a, m_b, m_c, m_d, A_i\}. \end{cases}$$
$$\Delta = \{and, or, xor\}.$$

Realization of all vector operations for logic calculating by using a single sequencer in Verilog environment and followed implementation in PLD chip gives the results:

Logic Block Utilization:

Number of 4 input LUTs: 400 out of 9,312 4%

Logic Distribution:

Number of occupied Slices: 200 out of 4,656 4%

Number of Slices only related logic: 200 out of 200 100 %

Total Number of 4 input LUTs: 400 out of 9,312 4%

Number of bonded IOBs: 88 out of 320 29%

Total equivalent gate count for design: 2400

## IV. CONCLUSION

Scientific novelty is presented by process-models for analyzing associative tables, based on use vector logic operations to solve the problems of information retrieval, diagnosis, pattern recognition and decision-making in the vector discrete Boolean space. The models are focused to achieving high speed of concurrent vector logical analysis of information that in the limit completely excludes the use of arithmetic operations, including for calculating the criteria of quality solution. A multiprocessor architecture for concurrent solving associative logic problems with a minimal set of vector logical operations, which provides high performance, minimal cost and low power consumption of LAMP, implemented in PLD chip, is proposed. Further research will be aimed at developing a prototype of multiprocessor and solving new practical problems by using the proposed algebra.